\documentclass[12pt,preprint]{aastex}

\begin{document}

\shorttitle{X-ray Source Contamination in Star-Forming Regions} 
\shortauthors{Getman et al.}

\slugcomment{Accepted for the ApJS Special Issue on the Chandra Carina Complex Project}

\title{Source Contamination in X-ray Studies of Star-Forming Regions: Application to the {\it Chandra} Carina Complex Project}

\author{Konstantin V.\ Getman\altaffilmark{1}, Patrick S.\ Broos\altaffilmark{1}, Eric D.\ Feigelson\altaffilmark{1}, Leisa K.\ Townsley\altaffilmark{1}, Matthew S.\ Povich\altaffilmark{1,2}, Gordon P.\ Garmire\altaffilmark{1}, Thierry Montmerle\altaffilmark{3}, Yoshinori Yonekura\altaffilmark{4,5}, Yasuo Fukui\altaffilmark{6} } 
\altaffiltext{1}{Department of Astronomy \& Astrophysics, 525 Davey Laboratory, Pennsylvania State University, University Park PA 16802, USA}
\altaffiltext{2}{NSF Astronomy and Astrophysics Postdoctoral Fellow}
\altaffiltext{3}{Institut d'Astrophysique de Paris, 98bis bd Arago, FR 75014 Paris, France}
\altaffiltext{4}{Department of Physical Science, Osaka Prefecture University, 1-1 Gakuen-cho, Sakai, Osaka 599-8531, Japan} 
\altaffiltext{5}{Center for Astronomy, Ibaraki University, 2-1-1 Bunkyo, Mito, Ibaraki 310-8512, Japan}
\altaffiltext{6}{Department of Astrophysics, Nagoya University, Furo-cho, Chikusa-ku, Nagoya 464-8602, Japan}

\begin{abstract}
We describe detailed simulations of X-ray-emitting populations to evaluate the levels of contamination by both Galactic and extragalactic X-ray sources unrelated to a star-forming region under study. For Galactic contaminations, we consider contribution from main-sequence stars and giants (not including cataclysmic variables and other classes of accretion-driven X-ray binary systems) as they make the dominant contribution at the position of the Carina Nebula. The simulations take into consideration a variety of technical factors involving a Galactic population synthesis model, stellar X-ray luminosity functions, $Chandra$ telescope response, source detection methodology, and possible spatial variations in the X-ray background and absorption through molecular clouds. When applied to the 1.42 square-degree field of the $Chandra$ Carina Complex Project (CCCP), the simulations predict $\sim5000$ contaminating sources (1 source per square arcminute of the survey), evenly distributed across the field. The results of the simulations are further employed in a companion CCCP study to assign membership probabilities to individual sources.
\end{abstract}


\keywords{Galaxy: disk; ISM: individual (Carina) - open
clusters and associations: individual (Carina) - stars:
formation - stars: pre-main sequence -
X-Rays: stars}

\section{Introduction \label{sec_introduction}}

Identifying stellar members associated with a star-forming region is important for studies of the stellar initial mass function (IMF), cluster dynamics, and star formation processes.  Stars with dusty protoplanetary disks are readily found in infrared (IR) surveys where disk emission is strong, but when the disks have dissipated, the stars are most easily identified by their strong X-ray flaring.  High spatial resolution X-ray observations of an active star forming region with the $Chandra$ X-ray Observatory will often reveal hundreds or thousands of pre-main sequence (PMS) stars, as well as OB stars \citep{Feigelson07}.   

An important challenge for membership studies is to distinguish young stars formed in a molecular cloud from various types of contaminant sources along the line-of-sight. In X-ray surveys of star-forming regions, stellar X-ray contaminants include main-sequence (MS) stars \citep[e.g.][]{Schmitt95, Schmitt97} and some types of giants \citep[e.g.][]{Pizzolato00,Gondoin05}. These foreground and background stellar contaminants have much less impact on surveys in X-ray images than IR images because magnetic activity in PMS stars is elevated $10^{1}-10^{4}$ above levels in older stars \citep[e.g.][]{Preibisch05}. For star forming regions located in the quadrant of the Galactic plane centered on the Galactic Center, contamination by cataclysmic variables should be also considered \citep[e.g.][]{Muno04}. X-ray surveys also suffer some contamination by extragalactic sources, mainly quasars and other active galactic nuclei (AGN) \citep[e.g.][]{Brandt05}. These can be seen through the Galactic Plane as faint, absorbed X-ray sources.

Careful simulations of contaminating populations can provide estimates for the number of contaminants, their spatial distribution, and some of their apparent X-ray and IR properties. These estimates can inform efforts to assign individual classifications to the detected sources \citep{Broos11b}. Such simulation studies have been performed for the Orion Nebula Cluster (ONC) and the Cepheus~B (Cep~B) star-forming regions \citep{Getman05,Getman06}. For the deep $Chandra$ observation of the ONC, $150-200$ AGNs, $15-20$ foreground stars, and zero background stars are predicted, constituting $\sim10-15$\% of the $>1600$ X-ray detected sources. For the shallow $Chandra$ observation of Cep~B, $20-30$ AGNs, $10-15$ foreground stars, and a few background stars are predicted, again $\sim10$\% of the $Chandra$ source population. 

The number and observed properties of contaminating populations depend on a number of factors: observatory sensitivity, source detection method, observation exposure time, survey field of view, Galactic direction, distance to the star forming region, and absorption through the local molecular cloud. Therefore, it is ill-advised simply to extrapolate the contamination results for the Cep~B and ONC $Chandra$ fields to substantially different X-ray observations.

The purpose of this study is to present our contaminant simulation method and its application to the $Chandra$ Carina Complex Project \citep[CCCP;][]{Townsley11a}. The simulation methods are refined from those described by \citet{Getman06}. The Carina Nebula (NGC~3372) is one of the richest nearby high-mass star-forming complexes, located $2.3$~kpc from the Sun towards the Galactic direction $(l,b)=(287^{\circ}.6, -0^{\circ}.8)$ \citep{Smith08}. The CCCP combines 22 spatially contiguous $Chandra$-ACIS-I pointings with nominal exposure of $60$~ksec covering $1.42$ square degrees around the chain of famous massive star-forming regions in Carina \citep{Townsley11a}. A relatively shallow molecular cloud \citep{Yonekura05} covers roughly 1/3 of the CCCP field (Figure \ref{fig_spatial}c). More than $14000$ CCCP point sources have been detected \citep{Broos11a}.

The simulations of Galactic stellar and extragalactic contaminating populations in CCCP are described in \S \ref{contaminants_main_section}. They show that $\sim5000$ contaminating sources are expected in the CCCP catalog. Basic properties of these simulated contaminating sources (spatial distribution, X-ray median energies, and $J$-band magnitudes) are presented in \S \ref{contamination_properties_section}. The comparison of the properties of the simulated contaminating sources and the real CCCP sources is considered, at a simplistic level, in \S \ref{qualitative_analysis_section}. Based on the results of this study, with additional information on X-ray variability, IR excess, and optical spectral types, individual membership classifications for the CCCP sources are derived by \citet{Broos11b}.

\section{Simulations of Cluster Contaminants \label{contaminants_main_section}}

\subsection{Extragalactic Sources \label{contaminants_agn_section}}

We evaluate the expected contamination by extragalactic X-ray sources using methods similar to the analyses of \citet{Getman05,Getman06}. Nominal (corresponding to the the high Galactic latitude fields with low Galactic absorbing column used for extragalactic studies) hard-band ($2-8$~keV) fluxes incident on {\it Chandra} are drawn from the $\log N - \log S$ distribution of extragalactic sources described by \citet{Moretti03} assuming a power law source spectrum consistent with flux-dependencies described by \citet{Brandt01}. More than 22600 extragalactic sources down to a nominal flux of $\log(F_X) = -16.7$~erg~cm$^{-2}$~s$^{-1}$ are predicted for the 1.42 deg$^{2}$ CCCP field. One hundred Monte Carlo simulations are constructed by placing $>22600$ artificial sources randomly across the CCCP field. Source photon indices are drawn from uniform distributions in the ranges  $-0.5 < \Gamma < 2$ and $1 < \Gamma < 2$ for nominal fluxes $\log(F_X) < -14.1$~erg~cm$^{-2}$~s$^{-1}$ and $\log(F_X) > -14.1$~erg~cm$^{-2}$~s$^{-1}$, respectively.  

The nominal flux of each simulated source is transformed to apparent (i.e., observed) flux, taking into consideration an extra absorbing column that is estimated as the sum of two components: a uniform H{\sc I} column density of $N_{H} \sim1.4 \times 10^{22}$ cm$^{-2}$ through the entire Galactic disk in the direction of Carina \citep{Dickey90}\footnote{This was obtained from NASA's HEASARC tool located at \url{http://heasarc.gsfc.nasa.gov/cgi-bin/Tools/w3nh/w3nh.pl.}} and spatially variable absorption through the Carina molecular clouds. The cloud component is derived from the velocity-integrated maps of CO line emission obtained with the NANTEN telescope by \citet{Yonekura05}.  The gas column density outside the dense molecular cores was estimated from the maps of $^{12}$CO emission intensity integrated over the velocity range of $-30$~to~$-10$~km~s$^{-1}$ ($I(^{12}$CO$)$~[K~km~s$^{-1}$]) using the relationship $N(H_2) = I(^{12}$CO$) \times 1.6 \times 10^{20}$ \citep{Yonekura05}. Adopting the gas-to-dust ratio of \citet{Vuong03}, these gas column densities correspond to visual extinctions up to $A_V \sim3$~mag in the molecular clouds.  The dense core column densities are evaluated from the C$^{18}$O measurements and correspond to $A_V \sim5-10$~mag \citep[Table~3 in][]{Yonekura05}.  

$Chandra$ source photon spectra were then simulated, and observed count rates and median energies were derived using the {\it fakeit} command of the XSPEC software package \citep{Arnaud96}.  This program samples photons from the AGN spectrum, calculating the soft X-ray absorption by the line-of-sight material, and convolving the results with the $Chandra$ mirror and ACIS detector response. Typical on-axis CCCP calibration redistribution matrix and auxiliary response calibration files were used \citep{Broos11a}.  After applying local background levels found in the CCCP data, we then remove very weak extragalactic sources that would have fallen below the CCCP source detection threshold\footnote{Position-dependent detection thresholds corresponding to the CCCP detection process were calculated using local background levels estimated from the CCCP data. The very sophisticated and innovative source detection technique used in CCCP \citep{Broos10a,Broos11a} can be approximated by thresholding the source significance statistic, $Signif$, which is calculated by the ACIS Extract (\url{http://www.astro.psu.edu/xray/docs/TARA/ae\_users\_guide.html}) package as the ratio of net counts to the uncertainty on that quantity. The distribution of source significance for all CCCP sources peaks at $Signif = 0.8$, has a very sharp decay below 0.8 encompassing only 3\% of sources, and has an exponential decay above 0.8. In our contamination simulations we thus choose the source detection threshold of $Signif = 0.8$.}.    

Our simulations predict that $\sim2500$ extragalactic sources should be detected in the CCCP observation.

\subsection{Galactic Sources \label{contaminants_stellar_section}}

It has been argued that the point source component of the hard ``Galactic ridge emission'', which stretches over 90 degrees in Galactic longitude and only 2 degrees in latitude around the Galactic Center, is mainly composed of cataclysmic variables \citep[e.g][]{Ebisawa05,Revnivtsev07,Hong09}. The current version of our simulation omits the distribution of hard ($\ge3$~keV) X-ray sources in the Galactic plane attributed to cataclysmic variables and other classes of accretion-driven X-ray binary systems. However, this population and its X-ray emission should be unimportant in the Carina star forming region because it is located 72 degrees away from the Galactic Center where the Galactic Ridge emission is weak \citep[see Figure 1 in][]{Revnivtsev06}.



We simulate the stellar contamination from main sequence and giant stars in the Galactic Plane in three stages.  First, Monte Carlo simulations of the Galactic stellar population expected within the CCCP field are constructed based on the `Besan\c{c}on model' of stellar populations \citep{Robin03}.  The simulations are computed by the Web service provided by the Besan\c{c}on group\footnote{\url{http://model.obs-besancon.fr/}.}.  Using models of Galactic structure, stellar mass function, and star formation history, the calculation gives the number distribution of stars along a chosen line-of-sight including individual spectral types, optical and near-infrared magnitudes, ages, distances, and source extinctions. An additional position-dependent absorption representing the Carina molecular clouds was added for stars behind Carina ($D > 2.3$~kpc). Besan\c{c}on model limitations include a simplistic assumption of uniform extinction throughout the Galactic Plane and omission of spiral arms. We run the Galactic population model with an unrealistically deep $V=28$~mag detection limit, as the faint stars are later omitted by X-ray selection.

Second, X-ray luminosities for each star obtained from a Besan\c{c}on model simulation are then estimated by Monte Carlo sampling of X-ray luminosity functions (XLFs) of MS stars measured in the solar neighborhood from {\it ROSAT} surveys \citep{Schmitt95, Schmitt97, Hunsch99}.  Based on these studies, we adopt the following lognormal XLFs with the following means and standard deviations in $\log L_X$ [erg~s$^{-1}$] in the {\it ROSAT} PSPC energy band ($0.1-2.4$~keV): $<28.20> \pm 0.79$ for MS F-type stars; $<27.94> \pm 1.02$ for MS G-type stars; $<27.86> \pm 0.65$ for MS K-type stars; and $<27.44> \pm 0.76$ for MS M-type stars.  For giant stars, we adopt $<28.91> \pm 0.86$ \citep{Pizzolato00}. X-ray spectral shapes were assumed to follow the stellar temperature-luminosity relations of \citet{Gudel98} and \citet{Pizzolato00} for MS and giant stars, respectively.  The adopted temperature-luminosity relations in the {\it ROSAT} ($0.1-2.4$~keV) band are: $\log(T)$~[MK]~$=-9.2 + 0.33\log(L_X)$~[erg~s$^{-1}$] for MS stars and $\log(T)$~[MK]~$=-5.1 + 0.2\log(L_X)$~[erg~s$^{-1}$] for giants.  

Third, given the X-ray luminosity in the {\it ROSAT} ($0.1-2.4$~keV) band, spectrum, distance, and line-of-sight absorption for each of the simulated stars, the count rate in the $Chandra$ ACIS detector in the ($0.5-8.0$~keV) band is obtained using the Portable Interactive Multi-Mission Simulator (PIMMS\footnote{\url{http://cxc.harvard.edu/ciao/ahelp/pimms.html}.})\footnote{At this stage of the simulations, it was not feasible to use XSPEC to derive count rates for our very large sample of simulated stars.}. The detectability of each simulated star is then evaluated by locally-defined thresholds of the source significance statistic, similar to our treatment of extragalactic contaminants (\S~\ref{contaminants_agn_section}).   For the stars that satisfy the detection threshold,  the XSPEC {\it fakeit} command was used to simulate the spectrum to derive X-ray median energies for individual stars.  

A typical run of the Besan\c{c}on Galactic disk population model of the 1.42 deg$^2$ CCCP exposure of the Carina Nebula region predicts $\sim200,000$ foreground (distance $< 2.3$~kpc) MS stars in the following mass distribution:  4,000 F stars, 10,000 G stars, 20,000 K stars, and 170,000 M stars.  About 1,000 foreground giants are present.  Behind the cloud at distances $2.3-5.0$~kpc, the model predicts $\sim930,000$ background stars: 30,000 F stars, 77,000 G stars, 180,000 K stars, 630,000 M stars, and 9,000 giants. The vast majority of these stars are undetectable in the CCCP X-ray exposures.  This is  expected from the appearance of star forming regions at distances around $1-3$~kpc in different wavebands: infrared images are typically dominated by dense populations of field stars, while X-ray images are dominated by PMS stars in the young clusters.  

A typical Monte Carlo run predicts $\sim1800$ foreground and $\sim900$ background field stars will be detected in the X-ray survey.  The foreground detections include: 150 F stars, 450 G stars, 150 K stars, 850 M stars, and 150 giants.  Approximately 20\%, 50\%, and 30\% of the simulated foreground detections have distances from the Sun in the ranges $< 0.5$~kpc, $0.5-1.5$~kpc, and $1.5-2.3$~kpc, respectively.  The background detections include: 100 F stars, 500 G stars, 10 K stars, 40 M stars, and 250 giants.  Approximately 35\%, 35\%, and 30\% of the simulated background detections have distances in the ranges $2.3-3.0$~kpc, $3.0-4.0$~kpc, and $4.0-5.0$~kpc, respectively.

Several caveats pertain to the simulation. First, the line of sight towards the Carina complex is tangent to the Sagittarius-Carina spiral arm, but the Besan\c{c}on model of the Galactic disk population (which our simulations rely on) lacks spiral arm modeling \citep{Robin03}. Second, X-ray luminosities of stars with ages $<1$~Gyr might be even higher than those of stars from the solar neighborhood \citep{Preibisch05}. Roughly $1/4$ of the simulated Besan\c{c}on stars have ages $<1$~Gyr. Third, due to computational reasons, the simulations of Galactic background stars were truncated at a distance of $5$~kpc. Due to these three factors, we might expect even larger stellar contamination populations. Fourth, on the other hand, a possible decrease in X-ray activity with mass and age in giant stars \citep{Pizzolato00} is ignored in our simulations. If one considers only the most plausible class of X-ray emitting giants, F and G giants with ages $<1$~Gyr, the expected number of detected giants in the CCCP field could be as small as $\sim$10 in the foreground and $\sim$10 in the background.

Finally, we note that many Galactic field stars will each produce $<3$ photons in the CCCP image, too faint for detection as a point source.  The undetected foreground stars produce a total of $\sim10,000$ X-ray photons in the soft ($0.5-2$~keV) band.  These photons will appear as a uniform diffuse X-ray component in the CCCP image. These stars likely contribute $\sim$1\%  of the observed $> 10^6$~count soft X-ray emission that pervades the CCCP field \citep{Townsley11b}.

\section{Properties of Cluster Contaminants \label{contamination_properties_section}}

Basic properties of the simulated contaminating populations $-$ spatial distributions, X-ray median energies, and $J$-band magnitudes $-$ are described here. These properties are employed by \citet{Broos11b} to establish Carina membership probabilities for each of the CCCP sources.

\subsection{Spatial Density and Distribution \label{spatial_contamination_subsection}}

Our simulations predict that roughly 5000 contaminants, or approximately 1 source arcmin$^{-2}$ in the 5112 arcmin$^2$ CCCP survey, will be detected in our CCCP observations.  About half of these are expected to be extragalactic sources, and half Galactic field stars.  With $>14000$ detected sources \citep{Broos11a}, contaminants are expected to comprise $\sim35$\% of the detected CCCP source population. For comparison, the source surface densities of the X-ray contaminants detected in the ONC and Cep~B observations (\S~\ref{sec_introduction}) are 0.8 and 0.2 source arcmin$^{-2}$, respectively. A variation among these numbers is a product of multiple competing factors such as, detection technique, exposure times, thicknesses of molecular clouds and distances to the clouds, Galactic coordinates.

\begin{figure}
\centering
 \includegraphics[angle=0.,width=5.5in]{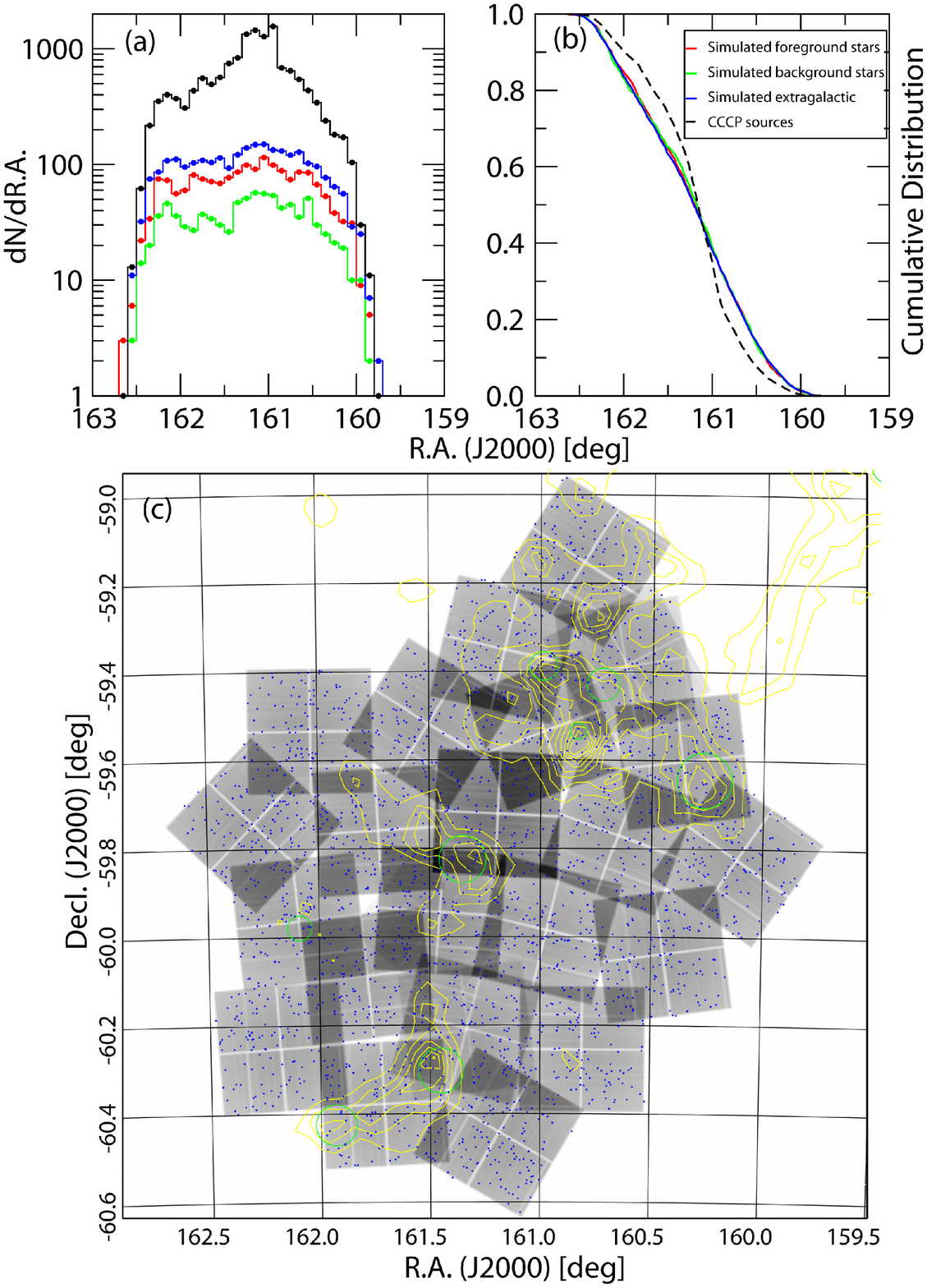}
\caption{Comparison of histogram (a) and cumulative (b) distributions in right ascension among the observed CCCP sources (black), simulated foreground stars (red), background stars (green), and extragalactic objects (blue). (c) Spatial distribution of simulated extragalactic objects (blue) superposed on the $Chandra$ Carina exposure map. The yellow contours show the $^{12}$CO emission and green contours indicate C$^{18}$O cores from \citet{Yonekura05}. \label{fig_spatial}}
\end{figure}

Across most of the Carina molecular cloud, the absorption is low, typically $A_V < 3$~mag, and reaches $\sim$10~mag only in a few small molecular cores \citep{Yonekura05}. Thus the spatial distributions of the simulated contaminants for all classes of the contaminants are relatively uniform across the field of view. We ignore small differences in the surface density of field stars between the northern (closer to the Galactic plane) and the southern portions of the CCCP field, predicted by the Besan\c{c}on model: $<2$\% and $<5$\% for the foreground and background stars, respectively. The spatial distribution of the contaminants is very different from the clustered distribution of the observed CCCP sources, as shown in maps of the source distributions \citep{Townsley11a, Feigelson11}.  The difference can be quantitatively seen in projection along the right ascension axis (Figure \ref{fig_spatial}). The lower numbers of simulated contaminating sources at the both ends of the right ascension range, compared to that of the central part of the field (Figure \ref{fig_spatial}a), are simply due to the decreasing angular coverage of the CCCP observation near the east and west boundaries (Figure \ref{fig_spatial}c).

\subsection{X-ray Median Energy \label{me_contamination_subsection}}

\begin{figure}
\centering
 \includegraphics[angle=0.,width=6.5in]{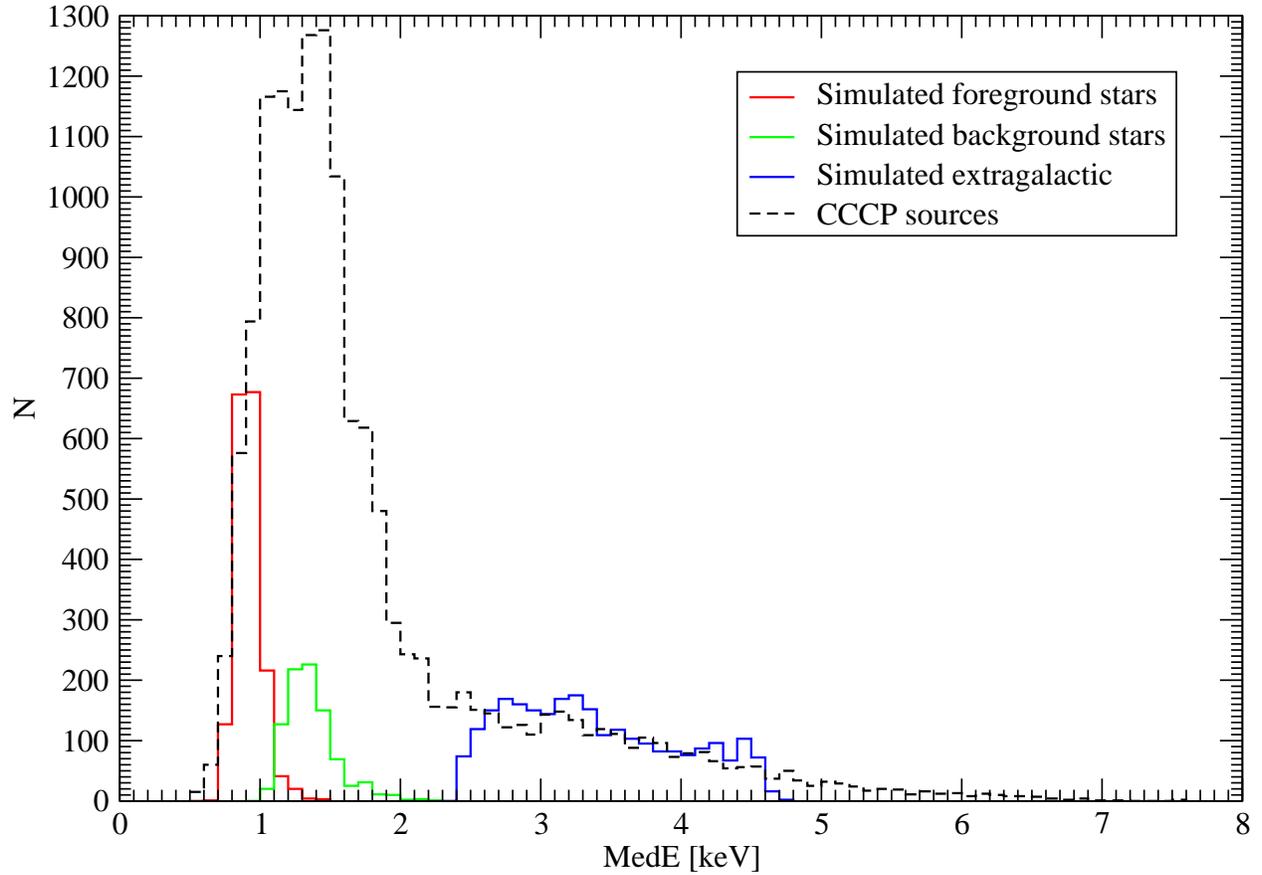}
\caption{Comparison of median energy distributions for the observed CCCP sources (black) and simulated foreground stars (red), background stars (green), and extragalactic objects (blue). \label{fig_me}}
\end{figure}

The X-ray median energy statistic, $MedE$, is the most robust among the measured photometric quantities used to describe the spectral shapes of weak X-ray sources. $MedE$ is most effective as a surrogate for the absorption column, but if the absorption is independently known, it can give a rough measure of plasma temperature \citep{Feigelson05,Getman10}.  From the simulated spectra of the Carina contaminants (\S~\ref{contaminants_main_section}), we obtain histograms of $MedE$ for each of the three simulated contaminant classes and compare them with the observed CCCP sources in Figure \ref{fig_me}. Notice that the observed CCCP sources are mixture of objects from all four classes: PMS stars, foreground stars, background stars, and AGNs.   A few important findings emerge. 
\begin{enumerate}

\item The three contaminant classes occupy mostly distinct ranges of $MedE$: 96\% of the simulated foreground stars have very soft X-ray spectra with $0.7 < MedE < 1.1$~keV range; 81\% of the simulated background stars are within the $1.1 < MedE < 1.5$~keV range; 94\% of the simulated extragalactic sources are within the $2.5 < MedE < 4.5$~keV range. 

\item The observed CCCP sources span a wide range of $MedE$, from 0.6 to 7~keV,  with a skewed distribution peaking at $1.4-1.5$~keV.  

\item The histograms of the observed CCCP sources and the simulated foreground stars have similar numbers of sources in the range $0.7 < MedE < 1$~keV, suggesting that nearly all the CCCP sources in this range are likely foreground stars.  This also implies that our simulation did not significantly underestimate the foreground population due to an unmodeled population of MS stars in the Carina spiral arm (\S~\ref{contaminants_stellar_section}).


\item The histograms of the observed CCCP sources and the simulated AGNs have similar numbers of sources in the range $2.5 < MedE < 4.5$~keV, suggesting that nearly all the CCCP sources in this energy range can be AGNs. 

\item  The peak of the simulated background stars coincides with the peak of the observed CCCP sources, but the numbers of the CCCP and simulated background sources are drastically different.  It is clearly impossible to distinguish between background stars and Carina PMS members using $MedE$.  

\end{enumerate}

We see that the X-ray median energy is an important discriminant for distinguishing some contaminant populations.  It can be effective for foreground stars and extragalactic sources, but not background stars which have the same median energies as many Carina PMS stars.

\subsection{\textit{J}-band Magnitude \label{jmag_contamination_subsection}}

Younger PMS stars with protoplanetary disks (infrared Class 0-I-II systems) can be readily distinguished from Galactic field stars by their infrared excess.  But often, as in the Carina complex, the majority of stars associated with a star forming region are older, diskless (Class III) systems.  X-ray studies are particularly effective in detecting Class~III stars which have high levels of magnetic activity. While in some X-ray studies of star forming regions, IR color-color and/or color-magnitude diagrams can be employed to successfully distinguish between {\it diskless} PMS and older stars \citep[e.g.][]{Getman05,Getman06,Kuhn10}, the small source extinction range (typically only $A_V \sim 1-3$~mag for most parts of the Carina region) and the large range of distances (only with respect to color-magnitude diagrams) for the simulated contaminating stars (\S \ref{contaminants_stellar_section}) blunt this technique. Therefore, here we restrict our consideration to a simple near-IR photometric property, $J$-band magnitude. In addition to $J$-band magnitude, \citet{Broos11b} employ some near-IR and mid-IR colors to distinguish between {\it disk-bearing} PMS and other stars. Wider usage of IR colors and incorporation of IR-to-X-ray flux ratios (when applicable) will be considered in our future studies of other star forming regions.

\begin{figure}
\centering
 \includegraphics[angle=0.,width=5.0in]{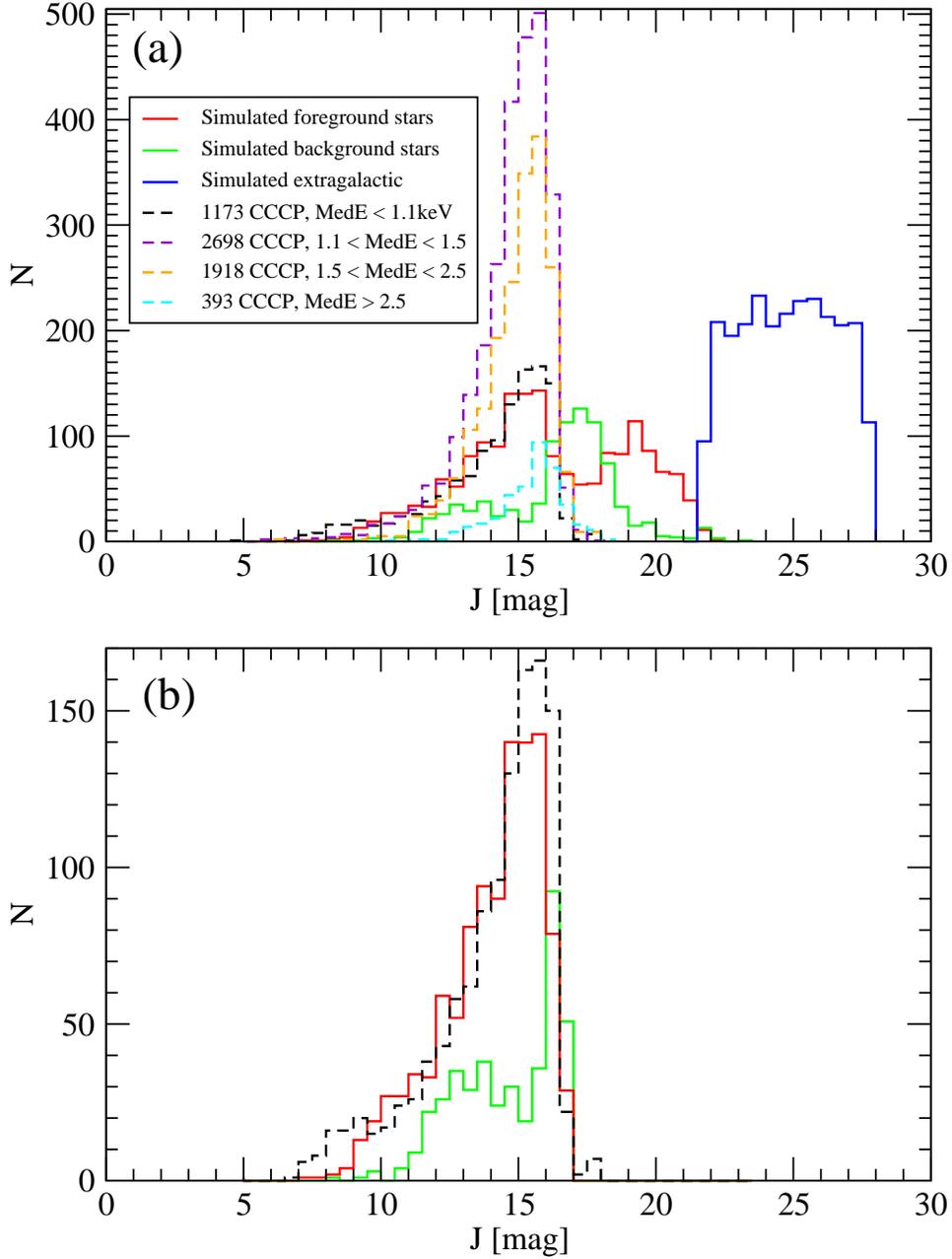}
\caption{Panel (a): comparison of $J$-band distributions for 6182 CCCP sources with available 2MASS counterparts (black, magenta, orange, cyan) and simulated foreground stars (red), background stars (green), and extragalactic objects (blue). The CCCP sources are stratified by median energy: $MedE < 1.1$~keV (black), $1.1 < MedE < 1.5$~keV (magenta), $1.5 < MedE < 2.5$~keV (orange), and $MedE > 2.5$~keV (cyan). Panel (b): Histograms of $J$-band magnitudes for simulated foreground (red) and background (green) stars multiplied by the 2MASS $J$-band detection completeness curve from \citet{Skrutskie06}. As a reference, the histogram of the CCCP sources with $MedE < 1.1$~keV (black) from panel (a) is also plotted here. \label{fig_jmag}}
\end{figure}

Figure \ref{fig_jmag} compares $J$-band magnitude histograms among the $6,182$ CCCP sources with available 2MASS counterparts within the entire CCCP field of view (dashed lines) and the three simulated contamination populations (solid lines). $J$-band magnitudes for the foreground and background field stars are obtained directly from the Besan\c{c}on model simulations; for the background stars, magnitudes are slightly increased for the absorption from the Carina molecular cloud. For the extragalactic contaminants, we produce an approximate distribution of $J$-band magnitudes using the relationship between the nominal (see \S \ref{contaminants_agn_section}) hard-band X-ray flux and $I$-band magnitude reported for the {\it Chandra} Deep Field North sources \citep[Figure~7 in][]{Alexander01}. The $I$-band magnitudes are transformed to $J$-band magnitudes following Figure~11 in Alexander et al., accounting for an average combined visual absorption of $A_V \sim10-12$~mag through the Galactic disk in the direction of Carina and through the Carina molecular clouds. For better compatibility with the CCCP histograms, we also present (Figure \ref{fig_jmag}b) the histograms of $J$-band magnitudes for simulated foreground and background stars multiplied by the ``generic'' 2MASS $J$-band detection completeness curve from \citet{Skrutskie06}\footnote{See also the 2MASS $J$-band detection completeness curves in the 2MASS All Sky Data Release Documentation at \url{http://www.ipac.caltech.edu/2mass/releases/allsky/doc/sec6\_5a1.html}. Notice, however, that this ``generic'' 2MASS detection completeness curve does not account for the effects specific to the CCCP field, such as the source crowding and IR nebulosity.}.

Several results emerge from the $J$-band magnitude distributions. First, the histogram of the simulated foreground population has two peaks (red curve). The fainter peak from M-type field dwarfs, constituting $\sim40$\% of the simulated foreground population, lies below the 2MASS sensitivity limit of $J \sim 16.5-17$~mag.  A bimodal shape is indeed seen in the deeper HAWK-I $J$-band magnitude histogram of the CCCP sources with $MedE < 1.1$~keV \citep{Preibisch11}. Second, for bright stars with $J \la 16$~mag, the $J$-band magnitude histogram of the simulated foreground population (red curve) agrees well both in number and shape with the magnitude distribution of CCCP sources with median energies $<1.1$~keV (black dashed curve).  This indicates, in agreement with the low-$MedE$ distributions discussed in \S~\ref{me_contamination_subsection}, that most of the observed CCCP sources in this sub-sample can be attributed to foreground stars. Third, roughly 60\% of the simulated stellar background population lies below the 2MASS sensitivity limit of $J \ga 16.5$~mag. Fourth, most of the simulated extragalactic objects lie below $J=21$~mag. The HAWK-I survey can detect some of these extragalactic objects near its detection limit \citep{Preibisch11}.

\section{Spatial Distributions Stratified by Median Energy and \textit{J}-band Magnitude \label{qualitative_analysis_section}}

\begin{figure}
\centering
 \includegraphics[angle=0.,width=5.0in]{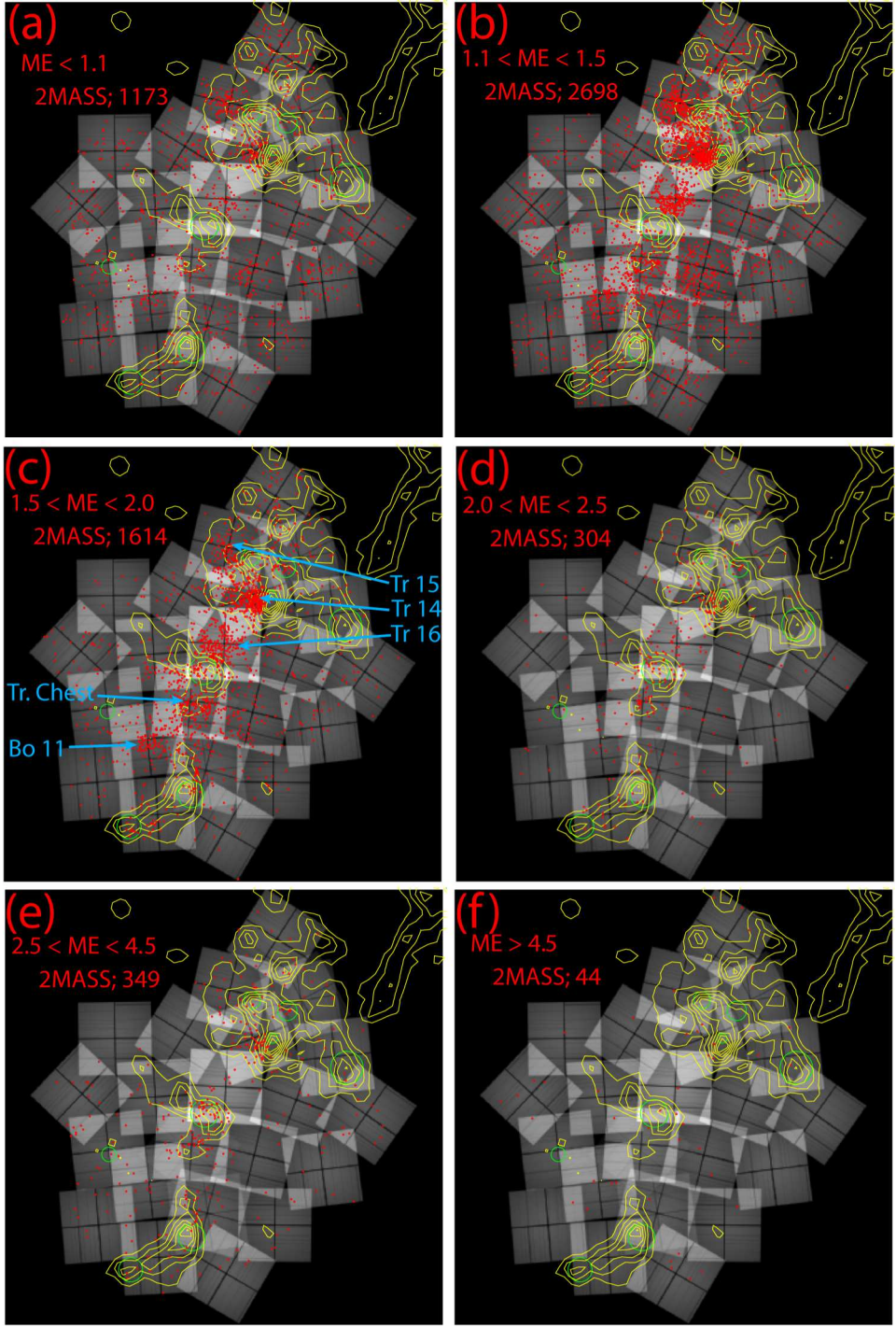}
\caption{Spatial distributions of all observed CCCP sources with 2MASS counterparts (red circles) superposed on the {\it Chandra} Carina exposure map. Distributions are stratified by X-ray median energy, a surrogate for line-of-sight absorption. Panels are stratified by X-ray median energies: (a) $MedE < 1.1$~keV; (b) $1.1< MedE <1.5$~keV; (c) $1.5< MedE <2.0$~keV; (d) $2.0< MedE <2.5$~keV; (e) $2.5< MedE <4.5$~keV; and (f) $MedE > 4.5$~keV.  The figure legends also provide numbers of plotted CCCP sources. The yellow contours show the $^{12}$CO emission and green contours indicate C$^{18}$O cores from \citet{Yonekura05}. \label{fig_aciswith2mass}}
\end{figure}

\begin{figure}
\centering
 \includegraphics[angle=0.,width=5.0in]{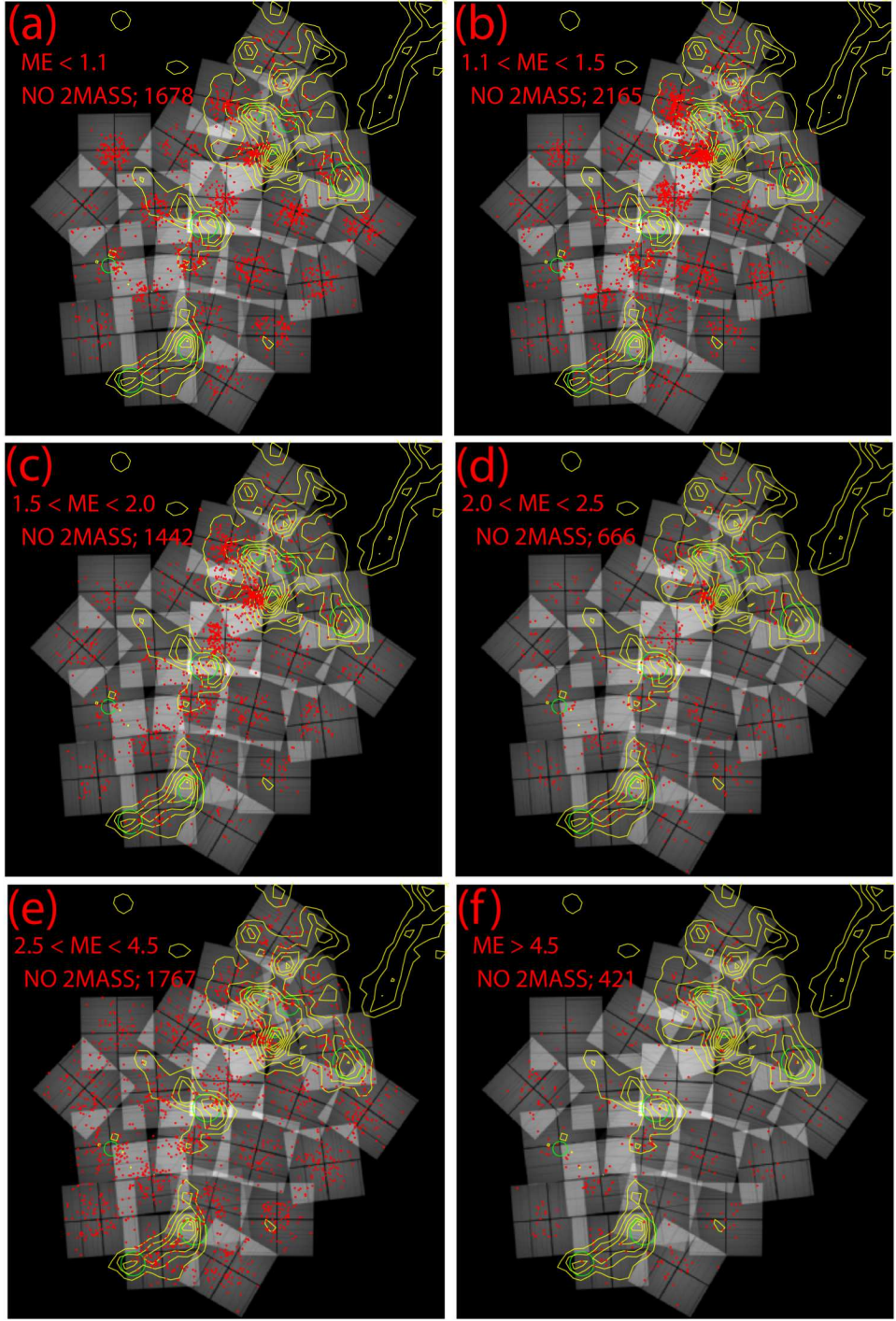}
\caption{Same as in Figure \ref{fig_aciswith2mass}, but for CCCP sources without 2MASS counterparts. \label{fig_acisno2mass}}
\end{figure}

Here we consider the spatial distribution of CCCP sources stratified by $MedE$ and by $J$-band magnitude that, as seen in \S~\ref{contamination_properties_section}, are often linked to contaminant classes.  We expect that Galactic and extragalactic contaminants will have random locations in the CCCP field, but many Carina members will lie in young stellar clusters.  The spatial distributions are shown in Figures \ref{fig_aciswith2mass}-\ref{fig_acisno2mass}, with  $MedE$ strata informed by Figure~\ref{fig_me}.  The contours show the distribution of molecular material from NANTEN CO maps \citep{Yonekura05}. The CCCP sources are further divided into two groups that have or lack 2MASS counterparts, giving two magnitude strata divided approximately at  $J \sim 16.5$~mag\footnote{A $\sim 3$\arcmin\/ circular `hole' in the distribution of 2MASS sources is seen in Figure \ref{fig_aciswith2mass}b due to saturation by $\eta$ Car in the 2MASS images.}. Interpretation of spatial distributions of CCCP sources is confused by the large variations in sensitivity due to overlapping exposures and off-axis angle \citep{Broos11a}. Our qualitative discussion here is designed to assess the role of contaminants.  A more quantitative treatment of spatial groupings from a spatially complete sub-sample of CCCP sources is given by \citet{Feigelson11}.

\subsection{Hard CCCP Sources\label{obscured_section}}

In this section we consider all CCCP sources with X-ray median energy $MedE > 2$~keV, equivalent to $A_V \ga 10$ mag \citep{Getman10}. In panels~$d-f$ of Figure~\ref{fig_aciswith2mass}, $70$\% of the $\sim700$ CCCP sources with 2MASS counterparts are spatially concentrated towards the edges of the Carina molecular clouds. Three prominent source groupings are seen:  members of the Treasure Chest cluster,  stars within the Trumpler~16 cluster, and stars at the interface between the rich Trumpler~14 cluster and C$^{18}$O core \#10. The latter region also has a dense grouping of $MSX$ mid-infrared sources. These groupings of absorbed X-ray sources are likely to be very young stellar objects. In contrast, out of $\sim2800$ hard CCCP sources without 2MASS counterparts (Figure~\ref{fig_acisno2mass}$d-f$), $\sim2000$ are evenly distributed across the CCCP field and are likely extragalactic candidates. This number is in satisfactory agreement with the number of simulated AGNs ($2500$ from \S \ref{contaminants_agn_section}).

\subsection{Soft CCCP Sources\label{unobscured_section}}

In this section we consider all CCCP sources with X-ray median energy $MedE < 2$~keV. Figure \ref{fig_aciswith2mass} presents all CCCP sources with available 2MASS counterparts. Panel~$a$ ($MedE<1.1$~keV) shows that only 40\% of the sources are spatially concentrated in the ridge of famous young stellar clusters in Carina (including Trumpler~15, Trumpler~14, Trumpler~16, Treasure Chest, and Bochum~11 \citep{Feigelson11})\footnote{The field of the cluster ridge is defined here as the lowest contour of the X-ray source density shown in Figure 1 of \citet{Feigelson11}.}, while the majority of the sources are relatively uniformly distributed across the CCCP field of view\footnote{In this section, source density estimates of evenly distributed populations are based on average source densities in the four ACIS-I fields outside the chain of the famous clusters. In \citet{Townsley11a} these four fields are labelled as E2, E4, SB1, and SB2.} (ignoring the ``egg-crate effect'', due to variations in sensitivity with off-axis angle \citep{Broos11a}). In contrast to panel~$a$, 70\% of the sources in panel~$b$ ($1.1< MedE <1.5$~keV) and 80\% of the sources in panel~$c$ ($1.5< MedE <2.0$~keV) are seen in projection against the chain of Carina clusters. Figure \ref{fig_acisno2mass} presents all CCCP sources without 2MASS counterparts. Panels~$a$ ($MedE <1.1$~keV), $b$ ($1.1< MedE <1.5$~keV), and $c$ ($1.5< MedE <2.0$~keV) show that approximately 80\%, 50\%, and 50\% of sources, respectively, are evenly distributed across the CCCP field.

Thus, based simply on their uniform spatial distributions, as many as $700$ ($+1400$) with $MedE<1.1$~keV, $800$ ($+1100$) with $1.1<MedE<1.5$~keV, and $300$ ($+700$) with $1.5<MedE<2.0$~keV sources with (without) 2MASS counterparts could be potential Galactic stellar contaminants. Further consideration of the expected $MedE$ and $J$-band properties for the simulated stellar contaminants (\S \ref{contamination_properties_section}) suggests that indeed, most of the $700+1400 = 2100$ sources with $MedE<1.1$~keV could be foreground stars, and roughly half of the $800+1100=1900$ sources with $1.1<MedE<1.5$~keV could be background stars. Therefore, this leaves us with at least $\sim300$ ($MedE<1.1$~keV),  $\sim1000$ ($1.1<MedE<1.5$~keV), and $\sim1000$ ($1.5<MedE<2.0$~keV) CCCP sources evenly distributed across the field that could be either additional contaminants unaccounted for by the simulations, and/or young stellar objects found outside the famous Carina clusters. In the former case, our underestimation of the Galactic stellar contamination population could be due to un-modeled young MS stars associated with Galactic spiral arm toward Carina, or background field stars with distances $>5$~kpc (\S \ref{contaminants_stellar_section}). In the latter case, we have an indication for the presence of a widely distributed population of young stars in the Carina complex.

\section{Summary}\label{summary_section}

This work describes simulations of X-ray-emitting source populations that will appear in high-resolution {\it Chandra X-ray Observatory} studies along the Galactic Plane.  The effort is particularly relevant to discriminating PMS stars in star forming regions from Galactic field star and extragalactic contamination. Our simulations account for two X-ray classes: extragalactic sources (primarily AGNs) seen through the Galactic disk and Galactic field stars (main sequence and giants from types F to M) distributed throughout the disk.  The simulations are based on a variety of technical considerations (\S~\ref{contaminants_main_section}):  use of a Galactic population synthesis model and measured X-ray luminosity functions to obtain field star X-ray source distributions; convolution of realistic contaminant X-ray intensity and spectral distributions through the telescope response;  consideration of spatially varying absorption across the field due to molecular clouds;  application of a realistic source detection method including threshold effects due to position-dependent point spread functions and X-ray background. The current version of the simulations omits contributions from accretion binary systems and will thereby underestimate the population at high $MedE$ for fields close to the Galactic Center.  

The results of the simulations are applied to the CCCP X-ray survey of the Carina star forming complex \citep{Townsley11a} to evaluate the levels of non-Carina contaminants among the $>14000$ CCCP point sources. The simulations predict that about 2500 AGNs (\S \ref{contaminants_agn_section}), 1800 foreground stars, and 900 background stars (\S \ref{contaminants_stellar_section}) will be detected. The simulated contaminants are expected to have uniform spatial distributions and exhibit significant differences in their distributions of the X-ray median energy and $J$-band magnitude (\S~\ref{contamination_properties_section}). These properties are thus useful for classifying individual CCCP sources as likely Carina members or contaminants \citep{Broos11b}.

Comparing properties of the simulated contaminating sources and the CCCP sources, at a simplistic level (\S \ref{qualitative_analysis_section}), we find that the number of hard CCCP sources evenly distributed across the field is in satisfactory agreement with our extragalactic contamination prediction, but the number of evenly-distributed soft CCCP sources is a factor of 2 higher than our Galactic contamination prediction (\S \ref{qualitative_analysis_section}). This discrepancy may be due either to unmodeled effects (e.g., stars beyond $D = 5$~kpc, an additional population of MS stars in the Carina spiral arm, or evolution of X-ray activity in older stars), or the presence of a widely distributed population of young stars in the Carina complex. The inference of distributed star formation, or the kinematic drifting of an older generation of Carina stars, is substantiated by more detailed studies of the CCCP sources lying outside the rich clusters \citep{Feigelson11, Preibisch11, Povich11b}. 

\acknowledgments We thank the anonymous referee for his time and many useful comments that improved this work. This work is supported by $Chandra$ GO grant SAO~GO8-9131X (L. Townsley, PI) and the $Chandra$ ACIS Team contract SV4-74018 (G. Garmire, PI), issued by the Chandra X-ray Center operated by the Smithsonian Astrophysical Observatory for and on behalf of NASA under contract NAS8-03060. M.S.P. is supported by an NSF Astronomy and Astrophysics Postdoctoral Fellowship under award AST-0901646. This publication makes use of data products from the Two Micron All Sky Survey (a joint project of the University of Massachusetts and the Infrared Processing and Analysis Center/California Institute of Technology, funded by NASA and NSF).

\facility{CXO (ACIS)}

\clearpage

\clearpage

\clearpage

\clearpage

\clearpage

\clearpage

\end{document}